\documentstyle [12pt]{article}
\begin{document}

\title{Application of $p$-adic analysis to models of spontaneous breaking
of the replica symmetry}
\author{V.A.Avetisov, A.H.Bikulov, S.V.Kozyrev}
\maketitle

\begin{abstract}
Methods of $p$-adic analysis are  applied
to the investigation of the spontaneous symmetry breaking in the models of
spin glasses.  A $p$-adic expression for the replica matrix is given and
moreover the replica matrix in the models of spontaneous breaking of the
replica symmetry in the simplest case  is expressed in the form of the
Vladimirov operator of $p$-adic fractional differentiation.  Also the
model of hierarchical diffusion (that was proposed to describe  relaxation
of spin glasses) investigated using $p$-adic analysis.
\end{abstract}

\section{Introduction}

Numerous works, for example
\cite{Parisi}, \cite{SpinGlass}, \cite{SpinGlass1} discuss
application of ultrametrics to investigation of spin glasses.
The most important example of ultrametric space is the field of
$p$-adic numbers, for introduction to $p$-adic analysis see
\cite{VVZ}.
In the present paper we apply the methods of $p$-adic analysis
to investigate the spontaneous symmetry breaking in the models of spin
glasses.  We obtain the following results:

1)\qquad
A $p$-adic expression for the replica matrix $Q_{ab}$ is found.
It has the form $Q_{ab}=q_k$  where $k=\log_p |l(a)-l(b)|_p$
where the notations is expressed below.
It is shown that the replica matrix in the Parisi form  \cite{Parisi}
in the models of spontaneous breaking of the replica symmetry
in the simplest case  have the form of the  Vladimirov operator of
$p$-adic fractional differentiation \cite{VVZ}.

2)\qquad
The model of hierarchical diffusion that was used in \cite{OS} to describe
relaxation of spin glasses
in our approach takes the form of the model of $p$-adic diffusion.
For instance, we reproduce the results of the paper  \cite{OS}
using the methods of $p$-adic analysis.

The models of spontaneous breaking of the replica symmetry are used
for investigations of spin glasses
\cite{Parisi}, \cite{SpinGlass}, \cite{SpinGlass1}.
The breaking of symmetry in such models is described by the replica
$n\times n$ matrix
${\bf Q}=\left(Q_{ab}\right)$ in the Parisi form \cite{Parisi}.
This matrix looks as follows. Let us consider the set of integer numbers
$m_i$, $i=1,\dots,N$, where  $m_i/m_{i-1}$
are integers for $i>1$ and $n/m_i$ are integers.
The matrix element of the replica matrix \cite{Parisi} is defined as
follows \begin{equation}\label{mat_el} Q_{aa}=0,\quad Q_{ab}=q_i,\qquad
\left[\frac{a}{m_{i}}\right]\ne \left[\frac{b}{m_{i}}\right];\quad
\left[\frac{a} {m_{i+1}} \right]= \left[\frac{b}{m_{i+1}} \right].
\end{equation}
Here $[\cdot]$ is the function of integer part
(we understand the integer part $[x]$ as follows: $[x]-1\le x\le [x]$
where $[x]$ is integer),
$q_i$ are some
(real) parameters.
An example of the matrix of this kind for
$m_i/m_{i-1}=2$ and $n=2^{N}$ have the form
\begin{equation}\label{matrix}
{\bf Q}=\left(
\begin{array}{ccccccccc}
{0}&q_{1}&q_{2}&q_{2}&q_{3}&q_{3}&q_{3}&q_{3}&\dots\\
q_{1}&{0}&q_{2}&q_{2}&q_{3}&q_{3}&q_{3}&q_{3}&\dots\\
q_{2}&q_{2}&{0}&q_{1}&q_{3}&q_{3}&q_{3}&q_{3}&\dots\\
q_{2}&q_{2}&q_{1}&{0}&q_{3}&q_{3}&q_{3}&q_{3}&\dots\\
q_{3}&q_{3}&q_{3}&q_{3}&{0}&q_{1}&q_{2}&q_{2}&\dots\\
q_{3}&q_{3}&q_{3}&q_{3}&q_{1}&{0}&q_{2}&q_{2}&\dots\\
q_{3}&q_{3}&q_{3}&q_{3}&q_{2}&q_{2}&{0}&q_{1}&\dots\\
q_{3}&q_{3}&q_{3}&q_{3}&q_{2}&q_{2}&q_{1}&{0}&\dots\\
\dots
\end{array}
\right)
\end{equation}
In the present paper we discuss the replica matrix
(\ref{matrix}) (more precisely, the generalization of this example
for the case of $p^N\times p^N$ matrices) using the language of
$p$-adic analysis. This allows to give the natural interpretation
for (\ref{matrix}) as the operator that can be diagonalized by
the $p$-adic Fourier transform. In particular this gives the spectrum
of the matrix  (\ref{matrix}).
In the limit of infinite breaking of the replica symmetry $N\to\infty$
the dimension $p^N$ of the replica matrix tends to infinity, but
the $p$-adic norm of the dimension $|p^N|_p=p^{-N}$ tends to zero.
The conjecture by Volovich \cite{volovich} is that
this phenomenon might explain the paradoxical fact that in the replica method
the dimension of the replica matrix in the limit of infinite breaking of
the replica symmetry tends to zero.

Let us make here a brief review of $p$-adic analysis.
The field $Q_p$ of $p$-adic numbers is the completion of the field of rational
numbers  $Q$ with respect to the $p$-adic norm on $Q$.
This norm is defined in the following way. An arbitrary rational number
$x$ can be written in the form $x=p^{\gamma}\frac{m}{n}$ with $m$ and $n$
not divisible by $p$. The $p$-adic norm of the rational number
$x=p^{\gamma}\frac{m}{n}$ is equal to $|x|_p=p^{-\gamma}$.

The most interesting property of the field of   $p$-adic numbers is
ultrametricity. This means that $Q_p$ obeys the strong triangle inequality
$$
|x+y|_p \le \max (|x|_p,|y|_p).
$$
We will consider disks in   $Q_p$ of the form
$\{x\in Q_p: |x-x_0|_p\le p^{-k}\}$.
For example, the ring $Z_p$ of integer $p$-adic numbers
is the disk
$\{x\in Q_p: |x|_p\le 1\}$
which is the completion of integers with the $p$-adic norm.
The main properties of disks in
arbitrary ultrametric space are the following:

{\bf 1.}\qquad
Every point of a disk is the center of this disk.

{\bf 2.}\qquad
Two disks either do not intersect or one of these disks
contains  the other.

The $p$-adic Fourier transform
$F$ of the function $f(x)$ is defined as follows
$$
F[f](\xi)=\widetilde{f}(\xi)=\int_{Q_p}\chi(\xi x)f(x)d\mu(x)
$$
Where $d\mu(x)$ is the Haar measure.
The inverse Fourier transform have the form
$$
F^{-1}[\widetilde{g}](x)=\int_{Q_p}\chi(-\xi x)\widetilde{g}(\xi)d\mu(\xi)
$$
Here $\chi(\xi x)=\exp(i\xi x)$ is the character of the field of
$p$-adic numbers.  For example, the Fourier transform of the indicator function
$\Omega(x)$ of the disk of radius 1 with center in zero (this is a function
that equals to 1 on the disk and to 0 outside the disk) is the function of the
same type:
$$
\widetilde{\Omega}(\xi)=\Omega(\xi)
$$

In the present paper we use the following Vladimirov operator
$D^{\alpha}_x$ of the fractional $p$-adic differentiation, that is
defined  \cite{VVZ}  as
\begin{equation}\label{diff}
D^{\alpha}_x
f(x)=F^{-1}\circ|\xi|_p^{\alpha}\circ F [f](x)=
\frac{p^{\alpha}-1}{1-p^{-1-\alpha}}
\int_{Q_p}\frac{f(x)-f(y)}{|x-y|_p^{1+\alpha}}d\mu(y)
\end{equation}
Here $F$ is the ($p$-adic) Fourier transform,
the second equality holds for $\alpha>0$.

For further reading on the subject of  $p$-adic analysis see
\cite{VVZ}.

\section{The replica matrix}

Let us describe the model of the replica symmetry breaking using
the language of $p$-adic analysis.  We will show that
the replica matrix ${\bf Q}=\left(Q_{ab}\right)$
can be considered as an operator in the space of functions on the finite set
consisting of $p^{N}$ points with the structure of the ring $p^{-N}Z/Z$.
The ring $p^{-N}Z/Z$ can be described as a set with the elements
$$
x=\sum_{j=1}^{N} x_j p^{-j},\qquad 0\le x_j\le p-1
$$
with natural operations of addition and multiplication up to modulus 1.
Let us consider the  $p$-adic norm on this ring
(the distance can take values  $0,p,\dots,p^{N}$).
We consider the following construction.
We introduce one-to-one correspondence
$$
l: 1,\dots,p^{N} \to p^{-N}Z/Z
$$
$$
l^{-1}: \sum_{j=1}^{N} x_j p^{-j} \mapsto 1+p^{-1}\sum_{j=1}^{N} x_j
p^{j}, \quad 0\le x_j \le p-1
$$
The formula (\ref{mat_el}) takes the form
\begin{equation}\label{mat_el_1}
Q_{aa}=0,\quad Q_{ab}=q_i,\qquad
\left[\frac{a}{p^{i-1}}\right]\ne \left[\frac{b}{p^{i-1}}\right], \quad
\left[\frac{a} {p^{i}} \right]= \left[\frac{b}{p^{i}} \right].
\end{equation}

Let us prove the following theorem.

{\bf Theorem.}
{\qquad\sl
The matrix element $Q_{ab}$
defined by (\ref{mat_el_1}) depends only on
$p$-adic distance between  $l(a)$ and $l(b)$:
$$
Q_{ab}=\rho(|l(a)-l(b)|_p),
$$
where $\rho(p^{k})=q_{k}$, $\rho(0)=0$.
}

{\it Proof}

The condition $\left[\frac{a} {p^{i}} \right]= \left[\frac{b}{p^{i}} \right]$
in our notions have a form
$$
\left[\frac{1+p^{-1}\sum_{j=1}^{N} a_j p^{j}} {p^{i}} \right]=
\left[\frac{1+p^{-1}\sum_{j=1}^{N} b_j p^{j}}{p^{i}} \right]
$$
This means that $a_j=b_j$ for $j>i$. The condition
$\left[\frac{a}{p^{i-1}}\right]\ne \left[\frac{b}{p^{i-1}}\right]$
means that $a_i\ne b_i$. But these two condition both mean that
$|l(a)-l(b)|_p=p^{-i}$.
We get that the matrix element of the replica matrix
$Q_{ab}$ depends only on the $p$-adic distance
$|l(a)-l(b)|_p$:  if $|l(a)-l(b)|_p$ equals to  $p^{-k}$
then $Q_{ab}=q_{k}$ and the statement of the theorem follows.

The replica matrix $\left(Q_{ab}\right)$ acts on functions on
$p^{-N}Z/Z$ as on vectors with matrix elements $f_b$
where $b=l(y)$, $b=1,\dots, p^N$.
The action of the replica matrix in the space of functions
on $p^{-N}Z/Z$ takes the form
\begin{equation}\label{action_discrete}
{\bf Q}f(x)=\int_{p^{-N}Z/Z}\rho(|x-y|_p)f(y) d\mu(y)
\end{equation}
where the measure $d\mu(y)$ of one point equals to 1
and $f_b=f(l(b))$
(because we can consider the index $b$ of the vector
as the index of the first column of the matrix $\left(Q_{ab}\right)$).

It is easy to see that operators of the form (\ref{action_discrete})
have the following properties:

{\bf 1.}\qquad
The operators (\ref{action_discrete}) commute with  operators of shift.
This means that the operators (\ref{action_discrete})
can be diagonalized by the Fourier transform
(in our case this is the discrete Fourier transform).

{\bf 2.}\qquad
The function $\rho$ depends on $p$-adic norm of argument.

{\bf 3.}\qquad
$\rho(0)=0$.

The language of $p$-adic analysis  allows us to describe
the natural generalization of the operator (\ref{action_discrete}).
This generalization have the form of operator
\begin{equation}\label{action}
{\bf Q} f(x)
= \int_{Q_p} \rho({|x-y|_p})f(y) d\mu(y)
\end{equation}
where the function $\rho$ obeys the properties {\bf 1.}-{\bf 2.}
(an analogue of the property {\bf 3.} will be considered later).
Here and further in the present paper we use the agreement that
we will use the same notions (without special comments) for
analogous values in the discrete and the continuous ($p$-adic) cases.

It is easy to see that the character $\chi(kx)$ is the generalized eigenvector
for the operator (\ref{action}), if $\rho({|x|_p})\in L^1(Q_p)$.
Thus the operator  (\ref{action}) can be diagonalized by the
$p$-adic Fourier transform  $F$:
$ {\bf Q} f(x)=F^{-1} \circ \gamma(\xi)  \circ F [f] (x) $.
From the property {\bf 2.} follows that the function $\gamma$
depends only on the $p$-adic norm of the argument:
$\gamma=\gamma(|\xi|_p)$.  Therefore we get
$$
{\bf Q}
f(x)=F^{-1} \circ \gamma(|\xi|_p)  \circ F [f] (x)
$$

\section{The model of hierarchical diffusion}

Let us reproduce now (partially) the results of the paper \cite{OS}
using the methods of $p$-adic analysis.
In the paper \cite{OS} relaxation of spin glasses
was described using the following
model of hierarchical diffusion.  Let us consider $2^{N}$
points (we will also consider more general case of $p^{N}$ points, $p>0$
is prime), separated by barriers of energy.
The barriers of energy have the following form.
Let us enumerate  the points by integer numbers starting from
$0$ to $2^{N}-1$ (analogously, from $0$ to $p^{N}-1$).
Let us consider the increasing sequence of the barriers of energy
(nonnegative numbers)
$0=\Delta_{0}<\Delta_{1}<\Delta_{2}<\dots<\Delta_{k}<\dots$.
We define the barriers of energy on the set of $p^{N}$ points
according to the following rule:
if $a-b$ is divisible by $p^{k}$ then the barrier between $a$-th and $b$-th
points equals to $\Delta_{k}$.

The hierarchical diffusion will be described by the ensemble of particles
that jump over the described above set of $p^{N}$ points.
Let us define the probability $q_{i}$ of transition (or jump) over the barrier
of energy $\Delta_i$ in the following way
$q_{i}=\exp(-\Delta_i)$, $i=1,2,\dots$.
Then the matrix of probabilities of transitions will be equal
(up to additive constant)
to the matrix ${\bf Q}$ of the form (\ref{matrix}).

We denote the density of particles at the $a$-th point as $f_a(t)$
and vector with elements that equal to densities at all points as
${\bf f}(t)$.  We define the dynamics of the model using the following
differential equation \cite{OS}:
\begin{equation}\label{equos}
\frac{d}{dt}{\bf f}(t) = ({\bf Q}-\lambda_0 {\bf
I}){\bf f}(t)
\end{equation}
where $N\times N$ matrix ${\bf Q}$ for $p=2$ have the form
(\ref{matrix}) of the replica matrix for the model of the replica
symmetry breaking, ${\bf I}$ is the unity matrix,
$\lambda_0$ is the eigenvalue of the matrix  ${\bf Q}$
that corresponds to the eigenvector with equal matrix elements.
This choice  of the transition probability matrix is defined by the law
of the conservation of number of particles
(that is an analogue of the property {\bf 3.}).

Application of the technique developed in the section 2 allows
us to write the equation (\ref{equos}) in the form
\begin{equation}\label{equos_int_discrete}
\frac{d}{dt}f(x,t)=\int_{p^{-N}Z/Z}(f(y,t)-f(x,t))\rho(|x-y|_p) d\mu(y)
\end{equation}
where $f_a(t)=f(l(a),t)$.
For example, for the considered above
$q_{i}=\exp(-\Delta_i)$, $i=1,2,\dots$ and for the linear dependence
of the barrier energy $\Delta_i=i(1+\alpha)\ln p$ on $i$
we get $\rho(|x|_p)=|x|_p^{-1-\alpha}$ and the equation
(\ref{equos_int_discrete}) gets the form
\begin{equation}\label{equos1}
\frac{d}{dt}f(x,t)=
\int_{p^{-N}Z/Z}
\frac{f(y,t)-f(x,t)}{|x-y|_p^{1+\alpha}} d\mu(y)
\end{equation}
At the right hand side of the equation (\ref{equos1})
we get the discretization of the Vladimirov operator
$D^{\alpha}_x$ (\ref{diff}) of the fractional  $p$-adic differentiation,
see \cite{VVZ}.

In the paper \cite{OS} the Cauchy problem for the equation (\ref{equos1})
with initial condition $f(x,0)=\delta_{x0}$ was investigated.
The dependence on time of the value $P_0(t)$
that in $p$-adic notions have the form
$$
P_0(t)=f(0,t)=\int_{p^{-N}Z/Z}\delta_{y0}f(y,t)d\mu(y)
$$
was found. In the present paper we will calculate the value $P_0(t)$
using the methods of $p$-adic analysis.

The $p$-adic generalization of the equation (\ref{equos_int_discrete})
have the following form
\begin{equation}\label{equos_int}
\frac{d}{dt}f(x,t)=\int_{Q_p}(f(y,t)-f(x,t))\rho(|x-y|_p) d\mu(y)
\end{equation}
Let us describe how to get the spectrum of the operator ${\bf D}$ at
the right hand side (or simply RHS) of
(\ref{equos_int}) (or the spectrum of times of relaxation for the model
of hierarchical diffusion \cite{OS}, describing spin glasses).
We will use the $p$-adic Fourier transform. It is easy to see that
the character $\chi(kx)$ is the generalized eigenfunction for the operator
at the RHS of (\ref{equos_int}) if
$\rho({|x|_p})\in L^1(Q_p\backslash U_{\epsilon})$,
where $U_{\epsilon}$ is the arbitrary neighborhood of 0,
or equivalently if $\forall k\in Z$ the series
$\sum_{i=k}^{\infty} |\rho(p^i)| p^{-i}$ converges.
For instance, 1 is the eigenfunction for the eigenvalue that equals to 0.
The proof is as follows:
$$
{\bf D} \chi(kx) =
\int_{Q_p}
(\chi(ky)-\chi(kx))\rho({|x-y|_p}) d\mu(y)=
$$
$$
=\chi(kx)
\int_{Q_p}
(\chi(k(y-x))-1)\rho({|x-y|_p}) d\mu(y)=
\chi(kx)
\int_{Q_p}
(\chi(ky)-1)\rho({|y|_p}) d\mu(y)
$$
To finish the proof we note that
$\chi(ky)$ is locally constant function that equals to 1
in some neighborhood of 0.
Using that the integral
$\int_{|y|_p\le p^i}\chi(ky) d\mu(y) = p^{i}$, if $|k|_p\le p^{-i}$
and equals to zero if $|k|_p > p^{-i}$, we get
\begin{equation}\label{spectrum}
{\bf D} \chi(kx) =
\left(-\left(1-p^{-1}\right)\sum_{p^{i}\le |k|_p}p^{-i} \rho(p^{i})
-\frac{p^{-1}}{|k|_p}\rho(|k|_p)\right)\chi(kx)
\end{equation}
This relation shows the correspondence between the spectrum of relaxation
times and the elements of the replica matrix in the form (\ref{matrix})
(here $q_i=\rho(p^{i})$).
The relation (\ref{spectrum}) reproduces the result obtained in
\cite{OS}, where more complicated technique was used.

Let us describe how to get the operator at the right hand side
of the equation (\ref{equos_int_discrete})
using the analogous  operator
(at the right hand side of the equation (\ref{equos_int})) on $Q_p$.
Consider the finite dimensional subspace $V_N\subset L^2(Q_p)$
of the following form. The subspace $V_N$ consists of functions
with zero average with support in $p^{-N}Z_p$
that are constants on disks of radius  1.
Therefore the dimension of the subspace $V_N$ equals to  $p^{N}-1$.
The operator at (\ref{equos_int}) maps this space into itself.
At the subspace $V_N$ the operator at the RHS of the equation
(\ref{equos_int}) takes the form
$$
{\bf D} f(x)=\int_{p^{-N}Z/Z}(f(y)-f(x))\rho(|x-y|_p) d\mu(y)
$$
that looks exactly like the operator at the RHS of the equation
(\ref{equos_int_discrete}).
But we will not get in this way the equation (\ref{equos_int_discrete})
because the operator at the RHS of (\ref{equos_int_discrete})
acts in the space of dimension that is larger by 1 than $V_N$.
This space can be obtained from the space $V_N$ by adding to $V_N$
the function that equals to 1 on the ball $p^{-N}Z_p$ (and to 0 outside).

Thus the model \cite{OS}
(up to the comments made above)
corresponds to the action of the operator of  $p$-adic
fractional differentiation at the subspace.

We investigate the following  $p$-adic generalization
of the model \cite{OS}.  Let us consider the Cauchy problem
for the $p$-adic generalization of the equation (\ref{equos1})
\begin{equation}\label{equ_final}
\frac{d}{dt}f(x,t)+ A
D_x^{\alpha} f(x,t)=0,
\end{equation}
that have the form of the equation of $p$-adic diffusion
that was investigated in the book \cite{VVZ}.
We take the initial equation for the equation
(\ref{equ_final}) of the form
\begin{equation}\label{init_cond}
f(x,0)=\delta(x),
\end{equation}
This means that we investigate the fundamental solution
of the equation (\ref{equ_final}).
After the Fourier transform the equation (\ref{equ_final}) takes the form
$$
\frac{d}{dt}\widetilde{f}(\xi,t)+
A |\xi|_p^{\alpha} \widetilde{f}(\xi,t)=0,
$$
The solution of this equation is
$\widetilde{f}(\xi,0)e^{-A |\xi|_p^{\alpha}t}$.
Because the Fourier transform of the $\delta$-function with support in zero
equals to 1, we get finally
$$
\widetilde{f}(\xi,t)=e^{-A |\xi|_p^{\alpha}t}
$$
\begin{equation}\label{solution}
f(x,t)=\int_{Q_p}\chi(-\xi x)e^{-A |\xi|_p^{\alpha}t}d\mu(\xi)
\end{equation}
As the $p$-adic generalization of $P_0(t)$ we consider the value
$$
P_0(t)=\int_{|x|_p\le 1}f(x,t) d\mu(x)=\int_{Q_p}\Omega(x)f(x,t) d\mu(x)
$$
(we use the same notion) that for the solution (\ref{solution})
takes the form
\begin{equation}\label{p-adic_answer}
\int_{Q_p}\Omega(\xi) e^{-A |\xi|_p^{\alpha}t}d\mu(\xi)=
\int_{|\xi|_p\le 1}e^{-A |\xi|_p^{\alpha}t}d\mu(\xi)=
\left(1-p^{-1}\right)\sum_{k=0}^{\infty}p^{-k}e^{-Ap^{-\alpha k}t}
\end{equation}
The answer we have got coincides with the answer obtained in \cite{OS}
for (\ref{equos_int_discrete}).
The value found in \cite{OS} (they use $p=2$) have a form
\begin{equation}\label{their_answer}
P_0(t)=\lim_{n\to\infty}
\left(2^{-n}+{1\over 2}\exp\left(\frac{R^{n+1}t}{1-R}\right)
\sum_{m=0}^{n-1}\exp\left(-m\ln 2 -\frac{2-R}{1-R}R^{m+1}t\right)\right)
\end{equation}
where $0<R<1$ is some constant. It is easy to see that
(\ref{p-adic_answer}) and (\ref{their_answer}) coincide for
$R=p^{-\alpha}$ and $A=\frac{2-R}{1-R}$.
We see that $p$-adic analysis allows us to investigate
the models of hierarchical diffusion using the simple and natural formalism.

\vspace{10mm}
{\bf Acknowledgements}

\vspace{3mm}
The authors are grateful to I.V.Volovich for discussion.
This work was partially supported by INTAS 96-0698,
RFFI 98-03-3353a, RFFI 96-15-97352  and RFFI 96-15-96131 grants.

\end{document}